# Beating the Best: Improving on AlphaFold2 at Protein Structure Prediction

Abbi Abdel-Rehim[1], Oghenejokpeme Orhobor[2], Hang, Lou[3], Hao, Ni[3,4], Ross D. King[1,4,5]

1. Department of Chemical Engineering and Biotechnology, University of Cambridge, UK
2. The National Institute of Agricultural Botany, UK.
3. Department of Mathematics, University College London, UK.
4. Alan Turing Institute, UK.
5. Department of Computer Science and Engineering, Chalmers University, Sweden.

Abstract

The goal of Protein Structure Prediction (PSP) problem is to predict a protein's 3D structure (confirmation) from its amino acid sequence. The problem has been a 'holy grail' of science since the Noble prize-winning work of Anfinsen demonstrated that protein conformation was determined by sequence. A recent and important step towards this goal was the development of AlphaFold2, currently the best PSP method. AlphaFold2 is probably the highest profile application of AI to science. Both AlphaFold2 and RoseTTAFold (another impressive PSP method) have been published and placed in the public domain (code & models). Stacking is a form of ensemble machine learning ML in which multiple baseline models are first learnt, then a meta-model is learnt using the outputs of the baseline level model to form a model that outperforms the base models. Stacking has been successful in many applications. We developed the ARStack PSP method by stacking AlphaFold2 and RoseTTAFold. ARStack significantly outperforms AlphaFold2. We rigorously demonstrate this using two sets of non-homologous proteins, and a test set of protein structures published after that of AlphaFold2 and RoseTTAFold. As more high quality prediction methods are published it is likely that ensemble methods will increasingly outperform any single method.

1 Introduction

The success of AlphaFold2 in the annual CASP competition for predicting proteins from sequence (CASP14) made world-wide headlines: 'One of the Biggest Problems in Biology Has Finally Been Solved' Scientific American, 'It will change everything' Nature, 'The breakthrough of 2021' Science, the most important achievement in AI - ever' Forbes. A more sober assessment has recently been published by the 'structural biology community' which concluded that 'an average of 25% additional residues can be confidently modelled when compared with homology modelling [1]. Without doubt AlphaFold2 is the highest profile application of AI to science.

Protein sequence determines structure, and protein structure determines function. The Protein Structure Prediction (PSP) problem is to predict a protein's 3D structure (confirmation) from its amino acid sequence [2]. This has been a 'holy grail' of science since the Noble prize-winning work of Anfinsen showed that protein conformation was determined by sequence [3].

There are two broad approaches to determining structure from sequence. The most obvious is to simply simulate the folding progress, as we understand all the necessary physics (Shroedinger's equation). Despite a great deal of effort this approach this approach has had limited success [4,5]. The fundamental difficulty is that proteins take a long time to fold relative to the simulation time steps, and over time the approximations necessary to make the computation feasible introduces errors. In the future quantum computers may change this.

The alternative approach is to use machine learning (ML) to directly learn a mapping between sequence and structure. This idea is an old one [6,7]. Its success has improved with vastly increased computational power in the last 30 years (~1 x $10^6$), vastly increased numbers of sequence (~1 x $10^6$), more structures (~1 x $10^3$), and better ML methods that can exploit the increased power and data (e.g. deep learning).

The evaluation of PSP progress has been greatly enhanced by the series of Critical Assessment of Protein Structure Prediction (CASP) experiments [8,9]. In 2020 the CASP competition for predicting proteins from sequence (CASP14) saw remarkable improvement by two competitors, AlphaFold2 [10,11] and RoseTTAFold [12]. While AlpaFold2 was the superior predictor, RoseTTAFold seems to be better at predicting a significant minority of proteins. The succeeding CASP15 competition held in 2022 demonstrated further improvements in structure prediction capabilities. Importantly, many of these improvements were possible due of the adoption of predictors deployed in the preceeding CASP14, particularly AlphaFold2. While CASP15 submissions demonstrated a variety of approaches to improve predictions, none appeared to utilise the more straight-forward application of stacking.

Stacking is a form of ensemble ML. In ensemble ML multiple learning methods are combined to obtain better predictive performance than could be obtained from any of the constituent learning algorithms alone [13, 14]. Stacking has been very successful in practice [15]. In stacking multiple baseline models are first learnt, then a meta-model is learnt using the outputs of the baseline level model. Stacking starts with a single task $T_i$, represented by a set of $p$ attributes $X_i = (x_{i,1}, x_{i,2}, .. , x_{i,p})$, and a unique prediction attribute $y_i$. Then $m$ baseline models are trained using $m$ baseline ML methods $A_j$, $j=1, …,m$. $A_j(X_i) = y_{ij}$. Then a ML method $\Phi$ (possibly different from any $A_j$) learns the relationship between the latent features $y_{ij}$ and $y_i$. In our application of stacking. The attributes represent the amino acid sequence. Two baseline models have previously been trained ($T_1$ AlphaFold2 and $T_2$ RoseTTAFold). The unique prediction attribute represents the protein conformation. Each baseline model predicts conformation. We used a neural network ML method to form the stacking method ARstack that learns the relationship between AlphaFold2 and RoseTTAFold predictions and the true conformation.

## 2 Methods

### 2.1 Protein Datasets

To train and test ARStack we constructed three datasets of proteins with known confirmations and sequences. All the proteins had a solved 3D structure appearing in the RCSB database and had of at least 50 amino acids. Each protein in our dataset has a unique Uniprot accession number. Within these constraints, the proteins were randomly selected.

As homologous proteins have similar structures we grouped the proteins into homologous sets based on protein family as assigned by UniProtKB. Dataset 1 and 2 contain proteins of human origin. We chose human proteins from the homology sets because of their medical importance. We randomly selected a protein from each homologous set for Dataset 1 and two proteins from each set for Dataset 2. See Supplementary materials for complete lists of proteins included in each dataset.

| Datasets | Number of proteins |
| --- | --- |
| Family represented once (Dataset 1) | 360 |
| Family represented twice (Dataset 2) | 478 |
| Newer proteins (Dataset 3) | 112 |

Table 1. The three datasets used in this study.

Note that while proteins in Dataset 1 and 2 may have been part of the training data for both predictors, the third dataset contains proteins published after their release.

### 2.2 Prediction Methods

AlphaFold2 (v.2.0.1) and RoseTTAFold (v. 1.0.0) were downloaded from Github (https://github.com/RosettaCommons/RoseTTAFold and https://github.com/deepmind/alphafold/releases respectively). Both predictors were installed on the computing cluster (BERZELIUS) at the National Supercomputer Centre at Linköping University. The sequences for all selected proteins were obtained from Uniprot, and used as input for predicting protein structures. Multiple sequence alignments were prepared using hhsuite3 according to established pipelines described in the original papers [10,11]. AlphaFold2 pipeline generates 5 predictions per structure ranked by their pTM scores. The top ranked prediction was used as the sole AlphaFold2 contribution in this study. In contrast RoseTTAFold generates a single structure. All structures generated have been made available in a dedicated github repository [16].

## 2.3 ML Descriptors

A key to the successful application of ML is to use an appropriate representations. We based our approach on previous successful approached to predicting protein secondary structure. We used the Uniref90 Multiple Sequence Alignments served as the foundation for our descriptors. For each position in the protein chain, the fractions for all 20 amino acids (as well as 1 for gaps) present in the MSA were calculated and made into a tuple. For the 10 positions flanking the beginning and end of the polypeptide chain, vectors containing zeros were generated. For each position in the chain, the vector in question was concatenated with those of the previous 10 positions as well as the following 10 positions, placing the relevant position at the centre. The resultant 431 unit tuple served as the final representation for each position in the chain (Figure 1).

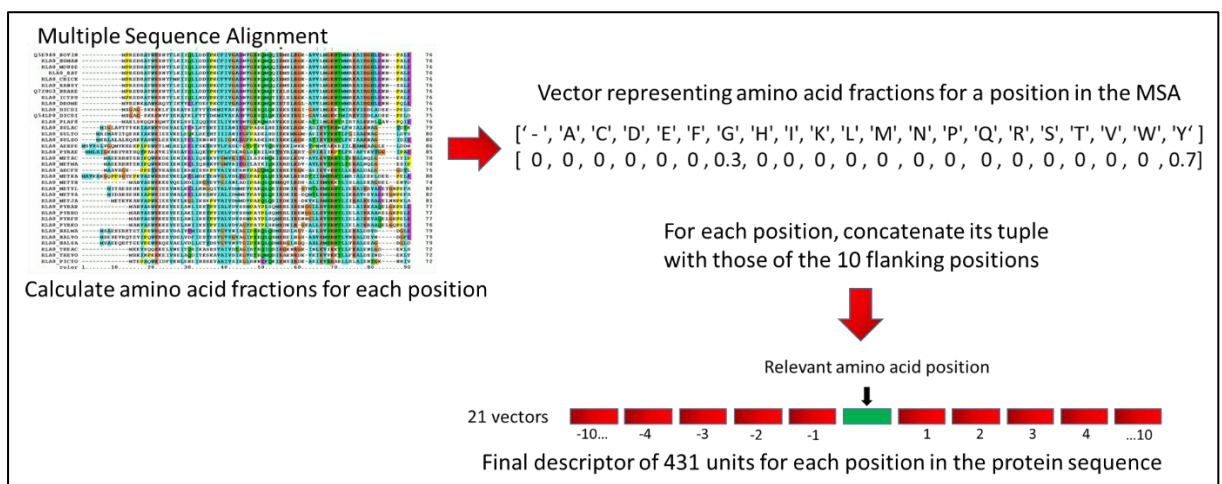

Figure 1. Deriving descriptors from an MSA. For each amino acid in a protein chain, the fractions of amino acids aligned to its position in the MSA was calculated. A 21 unit vector describing these fractions for each amino acid (and gaps) was generated. For each position in the protein, the relevant amino acid vector was concatenated with the 10 previous positions and the 10 following positions.

## 2.4 Target Values

We first aligned the sequences of the predicted structures to the experimental structures and performed any necessary truncations to generate a perfect sequence match. We then performed a superimposition using SVD (Singular value decomposition), implemented via the "SVDSuperimposer" module in BioPython [17]. After superimposing the predicted structures onto the experimental structures we extracted the transformed coordinates for the predicted structures. These transformed coordinates along with the original experimental coordinates were used for the operations necessary to acquire the target values. For each atom in the protein we calculated the vector from its predicted AlphaFold2 coordinates towards its RoseTTAFold coordinates. We posed the ML problem in the following way: predicting what position on this vector that is closest to the experimentally determined coordinates. We calculated the distance and direction from the AlphaFold2 coordinates to this point on the vector for each atom, these values served as target values (Figure 2).

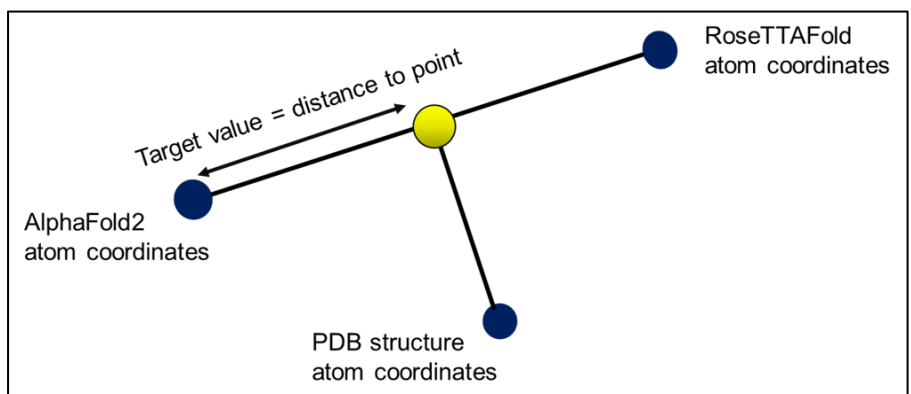

Figure 2. On the vector between the RoseTTAFold and AlphaFold2 coordinates, the distance and direction to the point that is closest to the coordinates of the experimental structure was calculated. These values were used as target values.

The target values calculated above will result in a column three times larger compared to that of the calculated descriptors due to each amino acid having three backbone atoms. However, as we are considering each backbone atom separately, no further alterations to the descriptors had to be made. A dataset with the calculated descriptors for each position in the polypeptide chain could be directly matched to the target values for the backbone atom in question.

The RoseTTAFold predictor produces structures with backbone atoms alone. Therefore, we only used these atoms in our study.

2.5 Machine Learning
We used Python Tensorflow and Keras libraries to create deep neural networks (DNN)[18]. Four layers were used, one input layer consisting of 431 neurons (matching the number of features), an output layer with a single neuron that outputs the predicted target value and three hidden layers with 200, 200 and 100 neurons respectively. A dropout value of 0.2 was used between the hidden layers. For the first two hidden layers, relu activation function was used, while the third hidden layer employed softmax. The optimisation function was set to stochastic gradient descent with a learning rate of 0.01, and the loss function to "mean squared error". We used three epochs and a batch size of 150 samples for the learning. Leave one (protein) out cross validation was used for evaluation.

3 Results
We applied stacking to combine AlphaFold2 and RoseTTAFold to develop ARStack. We found that ARStack significantly outperformed AlphaFold2 on all three datasets (Table 2, Figure 3). For Datasets 1 and 2 the results are highly significant. For the smaller Dataset 3 the result was significant at $P < 0.1$. *We conclude that the evidence strongly supports the conclusion that ARStack is a better predictor than AlphaFold2.*

| Dataset | **ARStack (n-wins)** | **AlphaFold2 (n-wins)** | % wins | p-value |
|---|---|---|---|---|
| 1 | 222 | 138 | 61.7 | $5.58 \times 10^{-6}$ |
| 2 | 284 | 194 | 59.4 | $2.23 \times 10^{-5}$ |
| 3 | 64 | 48 | 57.1 | 0.078 |

Table 2. Summary of results comparing ARStack and AlphaFold2 on three datasets. We used a one-tail binomial test to evaluate the statistical significance of the number of times ARStack beat AlphaFold2was significant. We selected this test for its robustness.

Dataset 1 contain one protein from each homology set. Using this form of stratified dataset is standard in prediction evaluation as if one knows the structure of one protein in a homologous set, then the others can be predicted using homology modelling. We found that our stacking approach worked better when addressing each of the backbone atoms separately. We compared the performance for the three backbone atoms using Dataset 1 (Table S1-S4). We found that the performance is similar for each of the three atom types, we therefore decided to report our results using the backbone atom 'CA'.

To demonstrate the benefit of using ARStack, we subtracted the AlphaFold2 RMSD values for each protein from those of ARStack (ΔRMSD), and plotted the differences for proteins across all three datasets (Figure 3). These plots clearly illustrate the distribution of proteins benefitting from either approach (positive values indicating better performance for ARStack and vice versa) and gives an idea of the absolute improvements across all proteins.

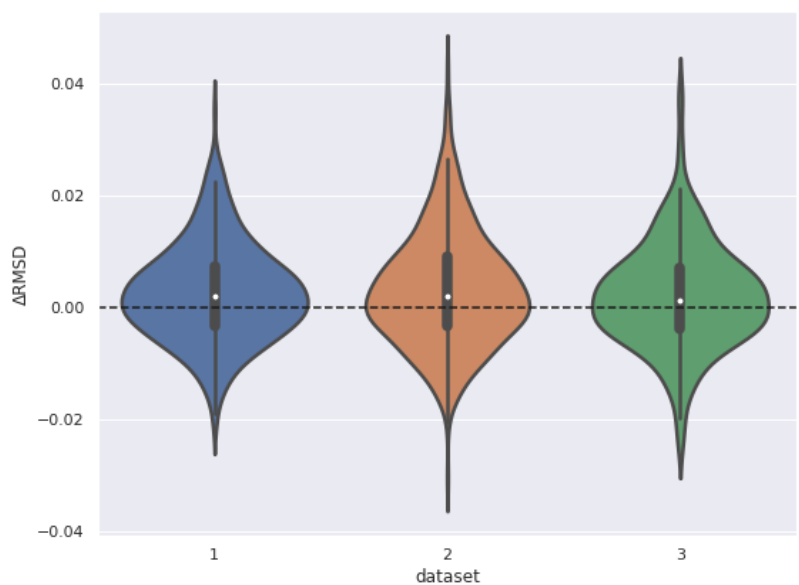

Figure 3. Comparing outcomes from ARStack and AlphaFold2 on 'CA' atoms across three datasets. Resultant RMSD calculations for ARStack were subtracted from those of AlphaFold2 to produce a difference (ΔRMSD), for all datasets. Where the differences are positive, The RMSD for the stacked approach is smaller than AlphaFold2 and vice versa.

Dataset 2 was designed to investigate the advantage of having homologous structures present in the training set. Despite each protein being unique, it would be likely that we could predict the left out protein with higher accuracy if a closely related protein is present in the training set. However, we found that this was not the case as can be seen in Table 2 and 3 (also Figure 3). The fraction of proteins where the stacked approach worked better is similar to that seen with Dataset 1.

Next, we analysed the differences in two separate groups based on which method that provides the better predictions, we call these groups "ARStack wins" and "AlphaFold2 wins" (Table 3). For instance, in cases where the stacked approach does better, the overall improvements (mean and median ΔRMSD) are higher for all atom types when compared to those that benefit more from using AlphaFold2. This demonstrates that the corrections gained by using the stacked approach outweighs those achieved when using AlphaFold2 alone.

| Δ RMSD - Å | **ARStack wins** | **AlphaFold2 wins** |
|---|---|---|
| Dataset 1 (mean) | 0.0085 | -0.0056 |
| Dataset 1 (median) | 0.0063 | -0.0046 |
| Dataset 2 (mean) | 0.0100 | -0.0063 |
| Dataset 2 (median) | 0.0078 | -0.0048 |
| Dataset 3 (mean) | 0.0077 | -0.0048 |
| Dataset 3 (median) | 0.0062 | -0.0036 |

Table 3. Proteins were separated into two groups based on which approach that does better. Within each of the two groups the difference between the two approaches were calculated (ΔRMSD) and their mean and median values reported.

Furthermore, we investigated the ΔRMSD for two different subsets for each dataset. The subsets differed in whether the AlphaFold2 predictions for the proteins were above or below 1 RMSD when compared to the experimental structures. The plot in Figure 4 clearly illustrates that ARStack provides more pronounced improvements compared to the Alphafold model on the proteins where the original model performs poorly (i.e. Alphafold RMSD is larger than 1 Å).

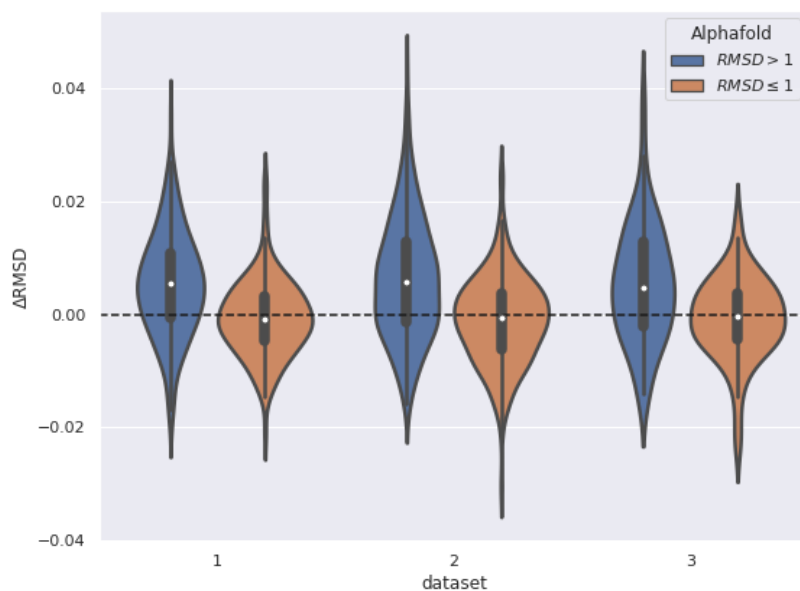

Figure 4. Comparing outcomes from ARStack and AlphaFold2 on 'CA' atoms across three datasets based on two cohorts of the AlphaFold2 prediction performances (AlphaFold2 RMSD>1 and AlphaFold2 RMSD< 1).

An additional observation is that despite the absolute improvements being higher for the stacked approach, when looking at the improvement as a percentage of the RMSD for the entire structure, AlphaFold2 appears to provide better performance (Table S3, S6 and S9). This is due to its performance being better in structures it does very well with to begin with (Table S4, S7 and S10). To be specific, the mean and median RMSD for the structures where AlphaFold2 is superior has a significantly lower RMSD overall, <50% compared to structures where the stacked perform better.

The statistically significant improvements of ARStack over AlphaFold2 in both Dataset 1 and Dataset 2 are based on cross-validation of sets of non-homologous proteins. This is the standard way to evaluate prediction methods in machine learning (Ross D. King, R.D., Orhobor, I.O., Taylor, C.C. (2021) Cross-validation is safe to use *Nature Machine Intelligence*. **3**, p. 276.). One limitation of these results is that neither AlphaFold2 nor RoseTTAFold have fully reported how they were trained. This means that there may be some statistical bias in the results, as AlphaFold2 or RoseTTAFold may have been trained on these proteins. We therefore prepared a third dataset, Dataset 3, where none of the protein structures could have been seen by either predictors (AlphaFold2 and RoseTTAFold) during their training, as they were placed in RCSB at a later date. The results of Dataset 3 shown in Table 2 (as well as Figure 3 and Table 3) are consistent with Dataset 1 and Dataset 2, confirming that ARStack outperforms AlphaFold2.

## 4 Discussion

We have demonstrated that a stacking neural network with a relatively basic architecture is able to improve the results by analysing multiple sequence alignments underpinning the performance of AlpahFold2 and RoseTTAFold. This is despite the fact that RoseTTAFold outperforms AlpahFold2 in only ~10% of cases (Figure S11).

To exemplify the ARStack improvements within a protein we use the structure of human cytoglobin H (Figure 5). This particular structure saw an improvement of 2.4% in its overall RMSD when using the stacked approach. In this structure it is clear that the majority of amino acids experience an improvement in the positioning of coordinates when using the stacked approach (blue). Apart from a handful of isolated amino acids, there is a significant part of one alpha helix that stands out as being better predicted by AlphaFold2 alone (red).

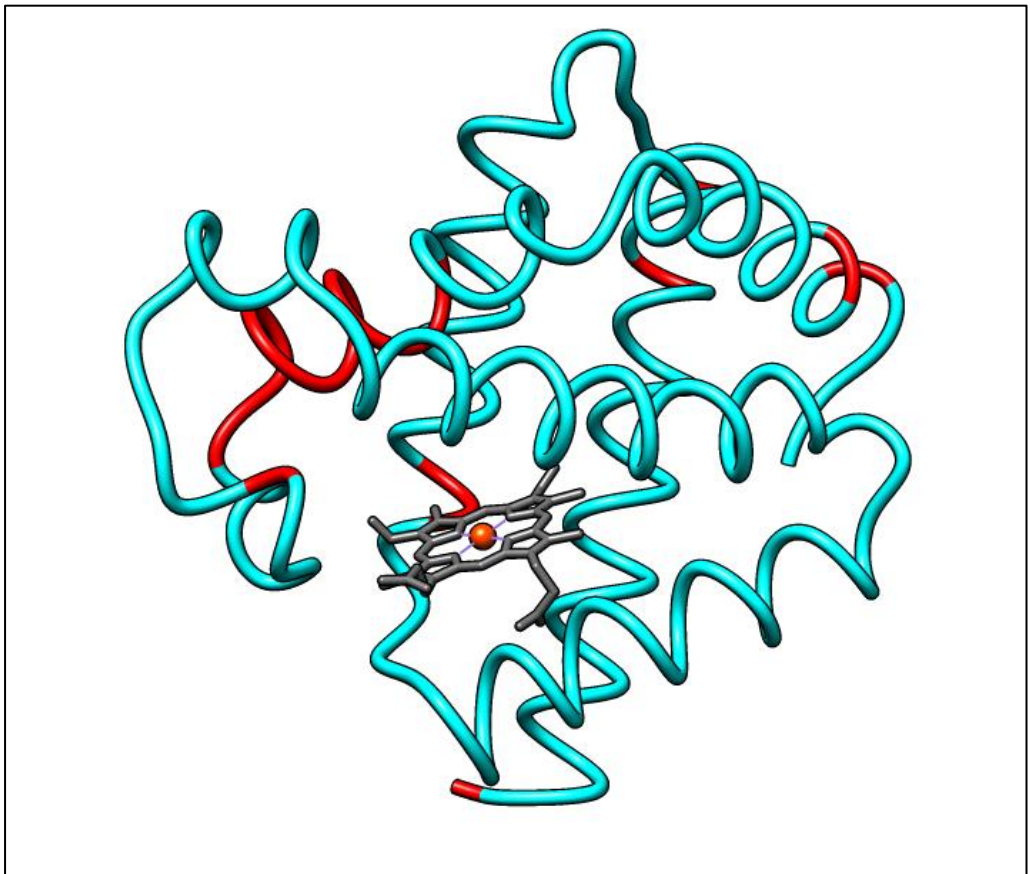

Figure 5. PDB structure "4B3W" of human Cytoglobin H benefited from the stacked approach.

Residues that were better predicted using the stacked approach are highlighted in blue as opposed to red (signifies that AlphaFold2 did better). The Heme group is shown in gray (Fe atom in orange).

We investigated whether there are any protein properties that are indicative of whether stacking provides an advantage. There seems to be two such properties amongst those we studied, namely molecular weight (MW) and flexibility. In smaller proteins stacking is more beneficial (Table 4). As flexibility is correlated with molecular weight and it is therefore expected that they should both reflect the same relationship to our predictors.

| Dataset 1 | Mean ΔRMSD(Å) | Mean MW | Median MW | Mean Flex | Median Flex |
|---|---|---|---|---|---|
| **ARStack** | 0.0168 | 35483 | 31312 | 304.1 | 261.5 |
| **AlphaFold2** | -0.0091 | 30826 | 22414 | 263.0 | 186.5 |

Table 4. Proteins in Dataset 1 were divided into two groups depending on which method that performs better (stacked or AlphaFold2). Within this group, proteins were sorted based on ΔRMSD. The mean and median molecular weight and protein flexibility were calculated for the top 50 predictions for each method. Mean ΔRMSD is also reported amongst this subset.

Taking a closer look at molecular weight, we can follow the benefit of using stacking on protein subsets across all three datasets (Figure 6). It quickly becomes apparent that the stacked approach performs most reliably on smaller proteins. For instance, if restricting oneself to only having used the stacked approach in cases below 10kDa, one would have noticed a benefit in 12 out of the 14 available structures in Dataset1 (Table 5). It is worth mentioning that the performance of RoseTTAFold on this group proteins were worse than AlphaFold2 in 13 out of 14 cases.

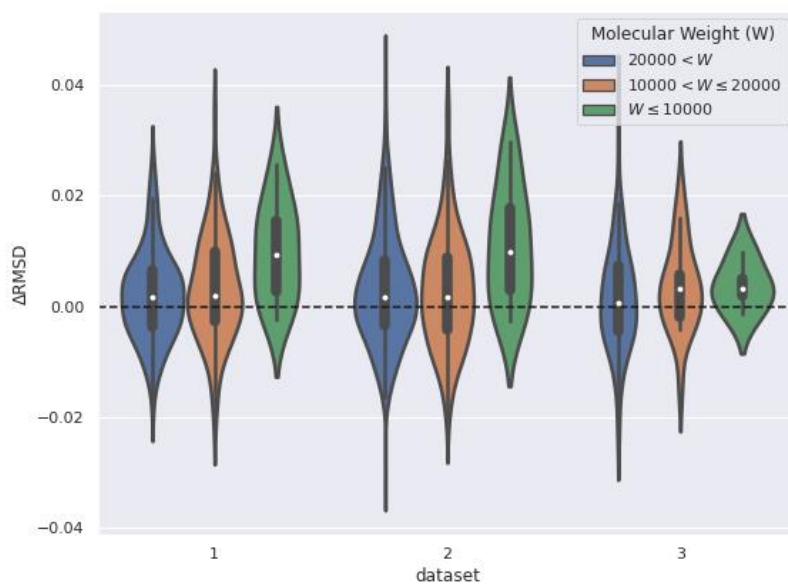

Figure 6. Correlation between stacking performance (in terms of ΔRMSD) and molecular weight across all three datasets.

| Molecular Weight | ARStack (n-wins) | AlphaFold2 (n-wins) | ARStack wins (%) | p-value |
|---|---|---|---|---|
| <10k | 12 | 2 | 85.7 | 0.0064 |
| 10k-20k | 59 | 35 | 62.8 | 0.0086 |
| All | 222 | 138 | 61.7 | $5.58*10^{-6}$ |

Table 5. Correlation between stacking performance and molecular weight. For each weight threshold, the number of proteins benefiting from either approach is declared. The percentage wins using the stacking approach is also shown. Statistical significance for the results is also shown for each threshold.

We performed this analysis on Dataset 3 as well. While ARStack outperforms AlphaFold2 in total wins, the significance of the results are different. As expected, the fractions of proteins benefitting from ARStack as well as the resultant p-values improve (Table 6) with smaller proteins.

| Molecular Weight | ARStack (n-wins) | AlphaFold2 (n-wins) | wins (%) | p-value |
|---|---|---|---|---|
| <20k | 18 | 7 | 72.0 | 0.022 |
| All | 64 | 48 | 57.1 | 0.078 |

Table 6. Correlation between stacking performance and molecular weight in Dataset 3. The number of proteins benefiting from either approach is declared for all proteins and proteins below 20kDa. The percentage wins for ARStack and the significance (p-value) of these results are also shown.

There is clearly room to expand and improve the descriptors as well as the learner in order to make further gains. One such opportunity is presented by a recently published protein folding predictor by Meta showing promising results for a large host of proteins [19].

CASP14 was a landmark event due to the unprecedented performance in predicting protein structure from sequence. The recently held CASP15 underscores this fact, showing that the most promising results were achieved when incorporating AlphaFold2 in some way. Sometimes this was done quite elaborately by combinations and/or alterations of components from established predictors. We could not find any team employing this more straight-forward application of stacking to combine predictions. Importantly, regardless of whether the new predictors surpasses the performance of ARStack shown in this paper, in principle ARStack should be able to perform even better when leveraging these improved methods performing at top level.

## 5 Conclusion

We have demonstrated that ARStack based on stacking AlphaFold2 and RoseTTAFold significantly outperforms AlphaFold2, the previous best PSP method. CASP14 was a landmark event due to the unprecedented performance in predicting protein structure from sequence by AlphaFold2. We argue that it is highly unlikely that any one single method will outperform all others on all proteins. Therefore, it is likely that the future of PSP will be based on ensemble prediction methods like ARStack.

## Acknowledgements.

This work was supported by the Wallenberg AI, Autonomous Systems and Software Program (WASP), Berzelius-2021-86, Data-Driven Life Science (DDLS) - SciLifeLab, and the UK Engineering and Physical Sciences Research Council (EPSRC) grant nos: EP/R022925/2 and EP/W004801/1. HN is supported by the EPSRC under the program grant EP/S026347/1 and the Alan Turing Institute under the EPSRC grant EP/N510129/1. HL is supported by University College London and the China Scholarship Council under the UCL-CSC scholarship (No. 201908060002).